\documentstyle[aps,epsf,floats]{revtex}  

\begin{document}        

\baselineskip 14pt

\title{Glueball Mass Spectrum from Supergravity\footnote{Based on two talks  
presented at 
DPF '99, UCLA, Los Angeles, CA, 5-9 January, 1999.}}
\author{Csaba Cs\'aki\footnote{Research fellow, Miller Institute
for Basic Research in Science.} and John Terning}
\address{Theoretical Physics Group \\
Ernest Orlando Lawrence Berkeley National Laboratory \\
University of California, Berkeley, CA 94720\\
{\rm and} \\
Department of Physics\\
University of California, Berkeley, CA 94720}

\maketitle

\vspace*{-5.5cm}
\begin{center}
 \hfill    LBNL-42987 \\
~{} \hfill UCB-PTH-99/08  \\
~{} \hfill hep-th/9903142\\
\end{center}
\vspace*{3.5cm}

\begin{abstract}       
We review the calculation of 
the spectrum of glueball masses in non-supersymmetric 
Yang-Mills theory using the conjectured duality between supergravity and
large $N$ gauge theories. The glueball masses are obtained by solving the
supergravity wave equations in a black hole geometry.
The glueball masses found this way are in
unexpected agreement with the available lattice data. We also show how
to use a modified version of the duality based on rotating branes to
calculate the glueball mass spectrum with some of the Kaluza-Klein 
states of the supergravity theory decoupled from the spectrum.

\end{abstract}

\section{Introduction}              

Maldacena's conjecture~\cite{Malda} relates ${\cal N}=4$ supersymmetric
$SU(N)$ gauge theories in the large $N$ limit to Type IIB string theory
on an AdS$_5\times {\bf S}^5$ background, where AdS$_5$ is a five dimensional
anti-de Sitter space. The metric of this space is given by
  \begin{equation}
 {ds^2 \over l_s^2 \sqrt{4 \pi g_s N}} = \rho^{-2} d\rho^2 +
\rho^2
\sum_{i=1}^4 dx_i^2 + d \Omega_5^2
\label{ads5}
\end{equation}
where $l_s$ is the string length related to the superstring
tension, $g_s$ is the string coupling constant
and $d\Omega_5$ is the line element on ${\bf S}^5$. The
$x_{1,2,3,4}$ directions in AdS$_5$ correspond  to ${\bf R}^4$
where the gauge theory lives. The gauge coupling constant $g_4$ of
the 4D theory is related to the string
coupling constant $g_s$ by $g_4^2 = g_s$.
In the 't Hooft limit
($N \rightarrow \infty$ with $g_4^2N = g_s N$ fixed),
the string coupling constant vanishes $g_s \rightarrow 0$.
Therefore we can study the 4D theory using the
first quantized string theory in the AdS space
(\ref{ads5}). Moreover if $g_s N \gg 1$, the curvature
of the AdS space is small and the string theory is approximated
by classical supergravity. Witten extended this proposal
to non-supersymmetric theories~\cite{witten}. In his setup 
supersymmetry is broken by heating up the ${\cal N}=4$ theory, which
corresponds to putting the four dimensional theory on a circle and
assigning anti-periodic boundary conditions to the fermions. In this case
the fermions will get a supersymmetry breaking mass term of the order
$T=1/2\pi R$, where $R$ is the radius of the compact coordinate and $T$ is
the corresponding temperature, while the scalars (not protected by
supersymmetry anymore) will get masses from loop corrections. Thus in
the $T\to \infty$ limit this should reproduce a pure (3 dimensional)
$SU(N)$ theory
in the large $N$ limit, which we will refer to as 
QCD$_3$. On the 
string theory side this corresponds to replacing the anti-de Sitter metric
by a Schwarzschild metric describing a black hole in the anti-de Sitter 
space. This metric is given by

\begin{equation}
  {ds^2 \over l_s^2 \sqrt{4 \pi g_s N}}
=
\left(\rho^2 - {b^4 \over \rho^2}
\right)^{-1} d\rho^2 +
\left(\rho^2 - {b^4 \over \rho^2}
\right) d \tau^2 + \rho^2
\sum_{i=1}^3 dx_i^2 + d\Omega_5^2,
\label{metric}
\end{equation}
where $\tau$ parameterizes the compactifying circle and
the $x_{1,2,3}$ direction corresponding to the ${\bf R}^3$
where QCD$_3$ lives. The horizon of this geometry is
located at $\rho=b$ with
\begin{equation}
    b = {1 \over 2R}=\pi T.
\label{horizonlocation}
\end{equation}
The supergravity approximation is valid for this theory when
the curvature of the space is small, thus when $g_s N\to \infty$. 
However, in order to obtain the pure gauge theory  
we have to take the
temperature to infinity. In order to keep the intrinsic
scale $g_3^2 N =g_4^2 N/R$ of the resulting theory at the scale of QCD, 
we simultaneously would need to take $g_4^2 N=g_s N \to 0$. Here $g_3$ is 
the dimensionful gauge coupling of  QCD$_3$.
This is exactly the opposite limit in which the supergravity approximation
is applicable! Thus as expected for any strong-weak duality, 
the weakly coupled classical supergravity theory and the QCD$_3$ theory
are valid in different limits of the 't Hooft coupling $g_4^2 N$. 

From the point of view of QCD$_3$, the radius $R$ of the
compactifying circle provides  the ultraviolet cutoff
scale. 
Therefore, with the currently available techniques,
the Maldacena-Witten conjecture
can only be used to study large $N$ QCD with a fixed ultraviolet
cutoff $R^{-1}$ in the strong ultraviolet coupling regime,
and hope that the results
one obtains this way are not very sensitive to removing the cutoff, that is
on going from one limit to the other.
Since the theory is non-supersymmetric, there is a priori no reason 
to believe that these two limits have anything to do with each other, since
for example there might very well be a phase transition when the
't Hooft coupling is decreased from the very large values where the 
supergravity description is valid to the small values where the theory 
should describe QCD$_3$. 
Nevertheless, Witten showed that the supergravity
theory correctly reproduces several of the qualitative features of a confining
3 dimensional pure gauge theory correctly \cite{witten}. In particular,
he showed that there is an area law in the Wilson loop and that there is a
mass gap in the spectrum, both of which are expected features of a 
confining gauge theory. Here we will address the question of whether
any of the quantitative features of the gauge theories are reproduced as 
well. In particular, we will calculate the glueball mass spectrum of the
theory, and find, that it is in reasonable agreement with recent lattice
simulations\cite{COOT}.

\section{The Glueball Spectrum in 3 Dimensions}

In this section we will show how to calculate the glueball spectrum of some
of the glueballs in the supergravity approximation in the 3 dimensional case. 
In the following we will
use the notation $J^{PC}$ for the glueballs, where $J$
is the glueball spin, and $P$, $C$
refer to the parity and charge conjugation quantum numbers respectively.
In the field theory, one can find operators that have the quantum numbers
corresponding to the given glueball states. For example, 
an operator with quantum numbers $0^{++}$ is given by
${\cal O}_4={\rm Tr} F^2$, or an operator with quantum numbers
$0^{--}$ is given by ${\cal O}_6=d^{abc} F_{\mu \alpha}^a F^{b\alpha \beta} 
F^c_{\beta \nu}$. According to the refinement of the
Maldacena conjecture given in \cite{ref}, one should find a supergravity
state corresponding to the chiral primary operators of the  original
${\cal N}=4$ conformal theory, which will couple to the supergravity 
states on the boundary of the AdS space. Assuming this coupling is 
maintained while heating the system, we can find the supergravity operators
coupling to ${\cal O}_4$ and ${\cal O}_6$. 
The dilaton and the R-R scalar of the supergravity theory
combine into a complex massless scalar 
field. Its real and imaginary parts
couple to the dimension 4 scalar operators ${\cal O}_4 = {\rm tr}~ F^2$ and
$\tilde {\cal O}_4 = {\rm tr} ~F\wedge F$. 
The NS-NS and R-R two-forms combine into a complex-valued
 antisymmetric field $A_{\mu\nu}$,
polarized along the ${\bf R}^4$. Its $({\rm AdS~ mass})^2=16$ and
thus one can  show that it couples to a dimension 6 two-form
operator of the ${\cal N}=4$ theory. This operator has been identified as
the operator ${\cal O}_6$
\cite{dastrivedi,flz}.
With this knowledge
we would like to calculate the actual glueball mass spectrum corresponding to 
these operators ${\cal O}_4$ and ${\cal O}_6$. In field theory, in order
to calculate the masses of these states one would need to evaluate the
correlators $\langle {\cal O}_4 (x)  {\cal O}_4 (y) \rangle =
\sum_i c_i e^{-m_i |x-y|}$, where the $m_i$'s are the glueball masses.
According to the refinement of the Maldacena conjecture \cite{ref},
this just amounts to solving the supergravity wave equations for
the fields that couple to these operators on the boundary. In the case
of the $0^{++}$ glueballs, we need to find the solutions of the
dilaton equations of motion of the form $\Phi =f(\rho ) e^{ikx}$. This
is because in the supergravity theory on AdS$_5\times {\bf S}^5$, the
Kaluza-Klein modes on the ${\bf S}^5$ can be classified according to the 
spherical harmonics of the ${\bf S}^5$, which form representations of
the isometry group $SO(6)$ (which is the R-symmetry group of the 
${\cal N}=4$ theory). When we put the theory at finite temperature, 
the states carrying non-trivial $SO(6)$ quantum numbers should eventually 
decouple from the spectrum, thus the glueballs should be identified with
the $SO(6)$ singlet states, which implies a solution of the form
$\Phi =f(\rho ) e^{ikx}$ for the dilaton as mentioned above.
Thus we will look for normalizable regular solutions to the dilaton
equation of motion which will give a discrete spectrum with the
glueball masses determined as $k_i^2=-M_i^2$.

In the supergravity description we have to solve the classical
equation of motion of the massless dilaton,
\begin{equation}
\label{dilaton}
\partial_{\mu} \left[ \sqrt{g} \partial_{\nu}\Phi
g^{\mu \nu} \right] =0 \ ,
\end{equation}
on the AdS$_5$ black hole background (\ref{metric}).
Plugging the ansatz $\Phi =f(\rho )e^{ikx}$ into this 
equation and using the metric of
(\ref{metric}) one obtains the following differential equation for
$f$:
\begin{equation}
\rho^{-1} {{d}\over{d \rho}} \left( \left(\rho^4 -
b^4 \right) \rho {{d f}\over{d \rho}} \right) - k^2 f = 0
\label{dilatondiff}
\end{equation}
Since the glueball mass $M^2$ is equal to $-k^2$, the task
is to solve this equation as an eigenvalue problem for $k^2$.
In the following we set $b=1$, so the masses are computed
in units of $b$. We need to find normalizable solutions to this equations
which are also regular at the horizon. For large $\rho$, the black hole
metric (\ref{metric}) asymptotically approaches
the AdS metric, and the
behavior of the solution for a $p$-form
for large $\rho$ takes the form  $\rho^{\lambda}$, where $\lambda$ is
determined from the mass $m$ of the supergravity field:
\begin{equation}
  m^2 = \lambda( \lambda + 4 - 2p)~.
\label{power}
\end{equation}
Indeed both (\ref{dilatondiff}) and (\ref{power}) give the
asymptotic forms $f \sim 1, \rho^{-4}$, and
only the later is a normalizable solution~\cite{witten}.
Changing variables to $f=\psi/\rho^4$ we have:
\begin{equation}
 \left( {\rho^2} - {\rho^6} \right) \,\psi^{\prime\prime}    +
  \left( 3\,{\rho^5} -7 \rho  \right) \,\psi^\prime
+\left( 16  + k^2  \rho^2 \right) \, \psi = 0
\end{equation}
For large $\rho$ this equation can be solved by series
solution with negative even powers:
\begin{equation}
\psi = \Sigma_{n=0}^\infty a_{2n} \rho^{-2n}
\label{asymp}
\end{equation}
Since the normalization is arbitrary we can set $a_0=1$.  The
first few coefficients are given by:
\begin{equation}
a_2= {{k^2}\over{12}}, \ \  a_4 = {{1}\over{2}} + {{k^4}\over{384}}, \ \ 
a_6 = {{7 k^2}\over{120}} + {{k^6}\over{23040}}.
\label{asymp2}
\end{equation}
For $n \ge 5$ the coefficients are given by the recursive relation:
\begin{equation}
(n^2+4n) a_n = k^2 a_{n-2} + n^2 a_{n-4} ~.
\end{equation}
Since the black hole geometry is regular at the horizon
$\rho=1$, $k^2$ has to be adjusted so that
$f$ is also regular at $\rho=1$ \cite{witten}. 
This can be done numerically in a 
simple fashion using 
a ``shooting" technique as follows.
For a given value of $k^2$ the equation is numerically integrated from
some sufficiently
large value of $\rho$ ($\rho \gg k^2$) by matching
$f(\rho)$ with the asymptotic
solution set by (\ref{asymp}) and (\ref{asymp2}).
The glueball mass $M$ is related to
the eigenvalues of $k^2$ by $M^2 = - k^2$ in units of $b^2$.
The results obtained this way, 
together with the results of the lattice simulations \cite{Teper97}
are displayed in Table~\ref{summary}. 
Since the lattice results are
in units of string tension, we normalize the supergravity
results  so that the lightest $0^{++}$ state agrees with
the lattice result.
 One should also expect a systematic error in addition to the  statistical
error
denoted in Table \ref{summary} for the lattice computations.
Similar numerical results have been obtained in \cite{jev}, 
while a WKB approximation for the eigenvalues of (\ref{dilatondiff}) has been 
obtained in \cite{Minahan}.
\begin{table}[htbp]
\centering
\caption{$0^{++}$ glueball masses in QCD$_3$
coupled to ${\rm tr}~F_{\mu \nu}
F^{\mu\nu}$. The lattice results are in units of the square root
of the string tension. The denoted
error in the lattice results is only the statistical one.\label{summary}}
\begin{tabular}{l|ccc}
state & lattice, $N=3$ & lattice, $N\rightarrow \infty$ &
supergravity  \\
 \hline
 $0^{++}$ & $4.329 \pm 0.041$ & $4.065 \pm 0.055$ & 4.07 ({\rm input}) \\
 $0^{++*}$ & $6.52 \pm 0.09$ & $6.18 \pm 0.13$ & 7.02 \\
 $0^{++**}$ & $8.23 \pm 0.17$ & $7.99 \pm 0.22$ & 9.92 \\
 $0^{++***}$ &  - & - & 12.80 \\
 $0^{++****}$ &  - & - & 15.67 \\
 $0^{++*****}$ & -  & - & 18.54 \\
\end{tabular}
\end{table}

The $0^{--}$ glueballs can be dealt with similarly by
considering  the two-form of the supergravity theory,
which couples to the operator ${\cal O}_6$.
The supergravity equation of motion
for the s-wave component of this field is given by 
\begin{equation}
{{3}\over{\sqrt{g}}} \partial_\mu\left[\sqrt{g} \, \partial_{[\mu^\prime}
A_{\mu_1^\prime
\mu_2^\prime]} \, g^{\mu^\prime \mu} g^{\mu_1^\prime \mu_1} g^{\mu_2^\prime
\mu_2}\right]
- 16  g^{\mu_1^\prime \mu_1} g^{\mu_2^\prime \mu_2}
A_{\mu_1^\prime \mu_2^\prime} = 0 ,
\end{equation}
where $[~~]$ denotes antisymmetrization with strength one.
For the pseudoscalar component of $A_{ij}$ the equation reduces to
\begin{equation}
 \rho\left({\rho^4}\, -  1\right) h''
 + \left( 3 + {\rho^4}\right)h'  -\left( {k^2}\,\rho\,
+16\,{\rho^3} \right)  h =0 ~,
\end{equation}
in units where $b=1$. This can be solved similarly as for the case of
the $0^{++}$ glueballs, and the results are displayed in Table \ref{3dtensor}.
Since the supergravity method and the lattice gauge theory
compute the glueball masses in different units,
one cannot compare the absolute values of the
lowest glueball mass obtained using these methods.
However it makes sense to compare the lowest glueball
masses of different quantum numbers.
Using Tables \ref{summary} and \ref{3dtensor}, we find that the supergravity
results
are in good agreement with the lattice gauge theory
computation \cite{Teper97}:
\begin{eqnarray}
&\left(\frac{M_{0^{--}}}{M_{0^{++}}}\right)_{{\rm supergravity}}&= 1.50 
\nonumber \\
&\left(\frac{M_{0^{--}}}{M_{0^{++}}}\right)_{{\rm lattice~~~~~}}& =
 1.45\pm 0.08
\end{eqnarray}

\begin{table}[htbp]
\centering
\caption{$0^{--}$ glueball masses in QCD$_3$
coupled to ${\cal O}_6$.
The lattice results are in units of square root of the string
tension. The normalization of the supergravity results
is the same as in Table \ref{summary}.\label{3dtensor}}
\begin{tabular}{l|ccc}
  state & lattice, $N=3$ & lattice, $N\rightarrow \infty$ &
supergravity  \\
 \hline
 $0^{--}$ &$6.48 \pm 0.09$ &$5.91 \pm 0.25$ & 6.10 \\
 $0^{--*}$ &$8.15 \pm 0.16$ & $7.63 \pm 0.37$ & 9.34 \\
 $0^{--**}$ & $9.81 \pm 0.26$ & $8.96 \pm 0.65$& 12.37 \\
 $0^{--***}$ & - & - &  15.33 \\
 $0^{--****}$ & - & -  & 18.26 \\
 $0^{--*****}$ & - &  - & 21.16 \\
\end{tabular}
\end{table}

One can see, that the glueball mass ratios obtained from the supergravity
calculation are in reasonable agreement with the lattice results, even though
as explained in the introduction these two calculations are in the 
opposite limits for  the 't Hooft coupling. Therefore, it is 
important to see, how the ratios are modified once corrections due to 
string theory are taken into account. The leading  string theory corrections
can be calculated by using the results of \cite{GKT}, who calculated the
first $\alpha'$ corrections to the AdS black-hole metric (\ref{metric}). 
The details of the calculation can be found in \cite{COOT}, here we just 
give the results for the $0^{++}$ state:
\begin{eqnarray}
M_{0^{++}}^2 &=& 11.59\times
 (1 -2.78  \zeta(3) \alpha'^3 ) \Lambda_{UV}^2 \nonumber \\
M_{0^{++*}}^2 &=& 34.53\times
 (1 -2.43  \zeta(3) \alpha'^3 ) \Lambda_{UV}^2 \nonumber \\
M_{0^{++**}}^2 &=& 68.98\times
 (1 - 2.28  \zeta(3) \alpha'^3 ) \Lambda_{UV}^2 \nonumber \\
M_{0^{++***}}^2 &=& 114.9\times
 (1 -2.23  \zeta(3) \alpha'^3 ) \Lambda_{UV}^2  \nonumber \\
M_{0^{++****}}^2 &=& 172.3\times
 (1 -2.21  \zeta(3) \alpha'^3 ) \Lambda_{UV}^2  \nonumber \\
M_{0^{++*****}}^2 &=& 241.2\times
 (1 -2.20  \zeta(3) \alpha'^3 ) \Lambda_{UV}^2 ~,
\label{truecorrection}
\end{eqnarray}
where $\Lambda_{UV}=\frac{1}{2R}$ and the correction to the horizon is
given by $b=(1-\frac{15}{8} \zeta (3) \alpha'^3)\frac{1}{2R}$. One can
see that the string theory corrections are somewhat uniform for the
different excited states of the $0^{++}$ glueball, and therefore 
one could hope that these corrections to the ratios of the 
glueball masses are small. However, it can be seen that this is probably a
too optimistic assumption, by considering the Kaluza-Klein partners of
the glueball states. As explained above, the glueball states do not 
carry quantum numbers under the $SO(6)$ isometry, and are also singlets
under the $U(1)$ symmetry corresponding to the compact direction $\tau$.
The Kaluza-Klein modes however do carry quantum numbers under 
$SO(6)\times U(1)$, and they do not correspond to any state in the
QCD theory, but rather they should decouple in the $R\to 0, g_4^2N\to 0$
limit from the spectrum. However, in the supergravity limit of finite
$R$, $g_4^2N \to \infty$ these states have masses comparable
to the light glueballs \cite{ORT}. This is simply a consequence of
the fact, that the masses of the fermions and scalars carrying the $SO(6)
\times U(1)$ quantum  numbers is of the order of the temperature $T$,
thus their bound states are expected to also have masses of the order of
the temperature. However, since the temperature is the only scale in 
the theory, and so 
this will also be the cutoff scale of the QCD theory, and thus
the mass scale for the glueballs.
In particular, the masses of the 
KK modes of the $0^{++}$ glueballs obtained from the dilaton equation 
by using the ansatz $\Phi= f(\rho ) e^{ikx} Y_l (\Omega_5)$ are given by
\cite{ORT}
\[
\begin{tabular}{c|cccc}
$l$ & 0 & 1 & 2 & 3 \\ \hline 
$M_l^2$ & 11.59 & 19.43 & 29.26 & 41.10 \\
${M_{l^{*}}}^2$ & 34.53 & 48.07 & 63.60 & 81.11 \\
${M_{l^{**}}}^2$ & 68.98 & 88.24 & 109.5 & 132.7 \end{tabular}
,\]
where we have displayed the unnormalized values of the masses of the
different KK modes.

One can explicitly see, that the masses of these KK modes are as 
expected of the same order as the masses of the glueball states. One 
might hope that even though the supergravity approximation of these masses 
is of the same order as for the glueballs, string theory corrections will
increase the masses of these states compared to the glueball states.
Unfortunately, at least the leading string theory corrections 
calculated in \cite{ORT,COOT} do not support this conclusion. The corrections
to the first few KK modes are

\begin{eqnarray}
M_{0}^2 &=& 11.59\times
 (1 -2.78  \zeta(3) \alpha'^3) \Lambda_{UV}^2 \nonumber \\
M_{1}^2 &=& 19.43\times
 (1 -2.73  \zeta(3) \alpha'^3) \Lambda_{UV}^2 \nonumber \\
M_{2}^2 &=& 29.26\times
 (1 - 2.74  \zeta(3) \alpha'^3) \Lambda_{UV}^2
\end{eqnarray}
Thus one can see that the masses of these KK modes in fact do need
large $\alpha'$ corrections to remove them from the spectrum of states. 
Then it is not clear why one would get large corrections to the masses
of the KK modes but not to the masses of the glueball states. This
situation is clearly unsatisfactory, therefore one may try to 
improve on it by introducing a different supergravity background, where 
some of these KK modes are automatically decoupled. We will consider this
possibility in the next section where we discuss the construction based
on rotating branes\cite{Russo,CORT,CRST}.

\section{The Glueball Spectrum in 4 Dimensions and the Construction Based
on Rotating Branes}

Results similar to the the ones presented in the previous section can be
obtained for the glueball mass spectrum in QCD$_4$ by starting 
from a slightly different construction where the M-theory 5-brane 
is wrapped on two circles \cite{witten}. The details of these results can be
found in \cite{COOT,HO}. Here we will review only the generalized
construction based on the rotating M5 brane with one angular momentum,
first constructed in \cite{Russo}, and explored in \cite{CORT}.
The metric for this background is given by
\begin{eqnarray}
ds^2_{\rm IIA}&=&{2\pi \lambda A \over  3u_0} u \Delta ^{1/2}\bigg[ 4u^2
\big( -dx_0^2+dx_1^2+dx_2^2+dx^2_3\big)
+ { 4A^2\over 9u_0^2} u^2 \ (1-{u_0^6\over u^6 \Delta }) d\theta_2^2
+ {4\ du^2  \over u^2 (1-{a^4\over u^4}-{u_0^6\over u^6 }) }
\nonumber \\
&+& d\theta^2+{\tilde{\Delta}\over \Delta} \sin^2\theta d\varphi^2
+{1\over \Delta } \cos^2\theta d\Omega_2^2
 -{4a^2 A u_0^2\over 3u^4\Delta } \sin^2\theta d\theta_2 d\varphi \bigg],
\label{pocho}
\end{eqnarray}
where $x_{0,1,2,3}$ are the coordinates along the  brane where the
gauge theory lives, $u$ is the ``radial" coordinate of the AdS space, while the
remaining four coordinates parameterize the angular variables of $S^4$, 
$a$ is the angular momentum parameter, and we have introduced
\begin{equation}
\Delta=1-{a^4\cos^2\theta \over u^4}\ ,\ \ \ \ \tilde{\Delta}=1-{a^4\over 
u^4} \ ,
\ \ \ \ A\equiv {u_0^4\over u_H^4-\frac{1}{3} a^4}\ , \ \ \ \ 
u_H^6-a^4 u_H^2-u_0^6=0\ .
\end{equation}
$u_H$ is the location of the horizon, and the dilaton background 
and the temperature of the field theory are  given by
\begin{eqnarray}
e^{2\phi }={8\pi\over 27} {A^3\lambda^3 u^3\Delta^{1/2}\over u_0^3} 
{1\over N^2}\ , \ \ \ \  R=(2\pi T_H)^{-1}={A\over 3u_0}\ .
\label{dilz}
\end{eqnarray}
Note, that in the limit when $a/u_0\gg 1$, the radius of compactification $R$
shrinks to zero, thus the KK modes on this compact direction are 
expected to decouple in this theory when we increase the angular momentum $a$.
In order to find the mass spectrum of the $0^{++}$ glueballs, we need
to again solve the dilaton equations of motion as a function of $a$. 
This can be done by plugging the background (\ref{pocho}) into the 
dilaton equation of motion 
\begin{equation}
\partial_{\mu} \left[ \sqrt{g} e^{-2 \Phi} 
g^{\mu \nu} \partial_{\nu} \Phi \right]=0.
\end{equation}
For a dilaton ansatz of the form $\Phi = f(u) e^{ikx}$ we obtain 
the differential equation
\begin{equation}
\partial_u \left[ u(u^6-a^4u^2-u_0^6) f'(u)\right]-k^2 u^3f(u)=0,
\end{equation}
which can be solved the same way as explained in the previous section,
where the eigenvalues are now a function of the angular momentum 
parameter $a$.
The results of this are summarized in Table \ref{tab:4ddila}. 
Note, that while some
of the KK modes decouple in the $a\to \infty$ limit, the 
$0^{++}$ glueball mass ratios change only very slightly, showing that the
supergravity predictions are robust for these ratios against the change
of the angular momentum parameter.

\begin{table}[htbp]
\centering
\caption{Masses of the first few $0^{++}$ glueballs in
QCD$_4$, in GeV,
from supergravity compared
to the available lattice results. The first column gives the lattice result
\protect\cite{Teper97,MorningstarPeardon,Peardon},
the second the supergravity result for $a=0$ while the third the
supergravity result in the $a\to \infty$ limit. The change
from $a=0$ to $a=\infty$ in the supergravity predictions is
tiny.
Note, that for the excited state the supergravity calculation came before
the lattice results.\label{tab:4ddila}}
\begin{tabular}{l|ccc}
state & lattice, $N=3$ &
supergravity  $a=0$ & supergravity $a\to \infty$ \\
 \hline
 $0^{++}$ & $1.61 \pm 0.15$   & 1.61 {\rm (input)} & 1.61 {\rm (input)} \\
 $0^{++*}$ &  $2.48 \pm 0.18 $  & 2.55 &  2.56 \\
 $0^{++**}$ &   - & 3.46  & 3.48 \\
 $0^{++***}$ &  -  & 4.36 &  4.40 \\
\end{tabular}
\end{table}

One can similarly calculate the mass ratios for the $0^{-+}$ glueballs,
by considering the equations of motion of the RR 1-form in the background
(\ref{pocho}), since on the D4 brane worldvolume this couples to
the operator ${\rm Tr}F\tilde{F}$. To find the glueball spectrum we 
have to solve the supergravity equation of motion of the RR 1-form
\begin{equation}
\partial_{\nu} \left[ \sqrt{g} g^{\mu\rho}g^{\nu\sigma}
(\partial_{\rho}A_{\sigma}-\partial_{\sigma}A_{\rho})\right]=0
\end{equation}
in the background (\ref{pocho}). 
Using the ansatz $A_{\theta_2} =f(u) e^{ikx}$
leads to the differential equation
\begin{equation}
(u^6-a^4u^2-u_0^6)\partial_u\left[ u^3(u^4-a^4)f'(u)\right]
-k^2u^5(u^4-a^4)f(u),
\end{equation}
which we solve using the same numerical methods as in the previous 
section.
The results are summarized in Table \ref{tab:0-+}. Note, that the
change in the $0^{-+}$ glueball mass is sizeable when going from $a=0$ to
$a\to \infty$, and is in the right direction as suggested by lattice results
\cite{MorningstarPeardon,Peardon}.

\begin{table}[htbp]
\centering
\caption{Masses of the first few $0^{-+}$ glueballs in
QCD$_4$, in GeV,
from supergravity compared
to the available lattice results. The first column gives the lattice result,
the second the supergravity result for $a=0$ while the third the
supergravity result in the $a\to \infty$ limit. Note that the change
from $a=0$ to $a=\infty$ in the supergravity predictions is 
of the order $\sim 25 \%$.\label{tab:0-+}}
\begin{tabular}{l|ccc}
state & lattice, $N=3$ &
supergravity  $a=0$ & supergravity $a\to \infty$ \\
 \hline
 $0^{-+}$ & 2.59 $\pm$0.13   & 2.00 & 2.56 \\
 $0^{-+*}$ &  3.64 $\pm$0.18   & 2.98 &  3.49 \\
 $0^{-+**}$ &   - & 3.91  & 4.40 \\
 $0^{-+***}$ &  -  & 4.83 &  5.30 \\
\end{tabular}
\end{table}

One can also calculate the masses of the different Kaluza-Klein modes
in the background of (\ref{pocho}). One finds, that as expected from the fact
that for $a\to \infty$ the compact circle shrinks to zero, the KK modes
on this compact circle decouple from the spectrum, leading to  a   
4 dimensional field theory in this limit. However, the KK modes 
of the sphere $S^4$ do not decouple from the spectrum 
even in the $a\to \infty$ limit. These conclusions remain unchanged even
in the case when one considers the theory with the maximal number of
angular momenta (which is two for the case of QCD$_4$) 
\cite{CRST,JorgeKostas}. 
In the limit when the angular momentum becomes large, $a/u_0\gg 1$,
the theory approaches a supersymmetric limit \cite{Russo,CRST}
since the supersymmetry breaking fermion masses get smaller
with increasing angular momentum \cite{CG}. Therefore, the limit of
increasing angular momentum on one hand does decouple some of the
KK modes which makes the theory four dimensional, but at the same time
reintroduces the light fermions into the spectrum \cite{CG}.

\section{Conclusions}
We have seen how the Witten extension of Maldacena's conjecture can be used
to study pure Yang-Mills theories in the large $N$ limit. 
These theories reproduce several of the qualitative features of QCD,
and one can also study the predictions for the glueball mass spectra.
One finds, that the supergravity calculations are 
in a reasonable agreement with the lattice results, even though they are 
obtained in the opposite limit
of the 't Hooft coupling. It would be very important to understand, whether
this unexpected agreement is purely a numerical coincidence or whether there
is any deeper reason behind it. 

\section*{Acknowledgements}
We thank Hirosi Ooguri, Yaron Oz, Jorge Russo and Konstadinos Sfetsos for
several collaborations, based on which this paper has been written. 
C. C.  is a research 
fellow of the Miller Institute for Basic Research in Science.
This work was supported in part
the U.S. Department of Energy under Contract DE-AC03-76SF00098, and in part
by the National Science Foundation under grant PHY-95-14797.

\end{document}